\documentstyle[preprint,aps,floats,tighten]{revtex}
\input{psfig.tex}
\begin{document}
\draft
\preprint{\vbox{\hbox{CU-TP-750} 
                \hbox{CAL-606}
                \hbox{astro-ph/9605100}
}}

\title{\vbox{\vskip 4cm DO WE KNOW THE GEOMETRY OF THE UNIVERSE?}}

\author{Marc
Kamionkowski\footnote{kamion@phys.columbia.edu}
and Nicolaos Toumbas}
\address{Department of Physics, Columbia University,
New York, New York~~10027~~USA}

\maketitle

\vskip2cm
\begin{abstract}
It is quite remarkable that seventy years after Hubble
discovered the expansion of the Universe, we still have no idea
in which of the three Friedmann-Robertson-Walker geometries we
live.  Most of the current literature has focussed on flat or open
models.  Here, we construct a viable model of the Universe which
has closed geometry even though the nonrelativistic-matter
density is less than critical.  Furthermore, in this model, the
cosmic microwave background could come from a causally-connected
region at the antipode of the closed Universe.  This model
illustrates that the geometry of the Universe is unconstrained
by current data.  We discuss observations which may reliably
determine the geometry of the Universe in the near
future.

(To appear in {\it Microwave Background
Anisotropies}, proceedings of the the XVIth Moriond Astrophysics
Meeting, Les Arcs, France, March 16--23, 1996.)
\end{abstract}

Remarkably, we have no idea in which of the three
Friedmann-Robertson-Walker (FRW) geometries we live, even
seventy years after the discovery of the expansion of the
Universe.  An open Universe accounts simply for a
nonrelativistic-matter density $\Omega_0$
that appears to be substantially less than unity.  Theoretical
arguments, such as the Dicke coincidence and inflation, favor a
flat Universe. Theorists have recently emphasized that the
Universe may be flat, even if $\Omega_0<1$, with a nonzero
cosmological constant.  There are also heuristic arguments for a
closed Universe.  However, given the current observations, it requires
some {\it chutzpah} to suggest that $\Omega_0>1$.  Here, we
describe a closed Universe with $\Omega_0<1$, which at low
redshifts is entirely indistinguishable from a standard open FRW
Universe with the same $\Omega_0$.  We also address how future
observations may determine the geometry of the Universe.

If matter with an equation of state $p=-\rho/3$
exists, then its energy density decreases as $a^{-2}$ with the
scale factor $a$ of the Universe, and thus
mimics a negative-curvature term in the Friedmann equation.
In this case, the Universe could be closed and still have
a nonrelativistic-matter density $\Omega_0<1$.$^{1-4)}$

The energy density contributed by a scalar field with a
uniform gradient-energy density would scale as $a^{-2}$,
but, such a scalar-field configuration would generically
collapse within a Hubble time.  Davis$^{1)}$
pointed out that
if there was a manifold of degenerate vacua with nontrivial
mappings into the three-sphere, then a texture---a topological
defect with uniform gradient-energy density---would be
stabilized provided that it was wound around a closed
Universe$^{1)}$.  Non-intersecting strings would also provide an
energy density that scales as $a^{-2}$.$^{3)}$

Moreover, if this energy density is chosen properly, the
observed cosmic microwave background (CMB) comes from a
causally-connected patch at the antipode of the closed
Universe$^{5)}$.  Although unusual, this model at least
looks remarkably like a standard open Universe at low redshifts,
even though the largest-scale structure differs dramatically.  

The Friedmann equation for a closed Universe
with nonrelativistic matter and matter (perhaps a stable
texture) with an equation of state $p=-\rho/3$ is
\begin{eqnarray}
     H^2 &\equiv& \left({\dot a \over a}\right)^2 = {8\pi G \over 3}
     \rho_m + {\gamma - 1 \over a^2} \nonumber \\
     &=&H_0^2 [ \Omega_0 (1+z)^3 +(1-\Omega_0)(1+z)^2]
     \equiv H_0^2 [E(z)]^2,
\label{friedmann}
\end{eqnarray}
where $H=\dot a/a$ is the Hubble parameter (and the dot denotes
derivative with respect to time), $z=(a_0/a)-1$ is the redshift,
$G$ is Newton's gravitational constant, $\rho_m$ is the density of
nonrelativistic matter, and $\gamma$ is a parameter that
quantifies the contribution of the energy density of the
texture.  The second line defines the function $E(z)$.
This is exactly the same as the Friedmann equation for an open
Universe with the same $\Omega_0$, so this closed Universe has
the same expansion dynamics.  At the current
epoch (denoted by the subscript ``0''),
\begin{equation}
     \Omega_0=1 + {1-\gamma \over a^2 H^2} = 1-\Omega_t +
     {1\over a_0^2 H_0^2},
\label{Omegaequation}
\end{equation}
where $\Omega_t=\gamma(a_0 H_0)^{-2}$ is the contribution of the
texture to closure density today.  So, $\Omega_0<1$ if
$\gamma>1$ even though the Universe is closed, and we require
that $\Omega_t + \Omega_0>1$.

If the metric of a closed Universe is written as
\begin{equation}
     ds^2= dt^2 - a^2(t) \left[ d\chi^2 + \sin^2 \chi (
     d\theta^2 + \sin^2 \theta d\phi^2) \right],
\end{equation}
then the polar-coordinate distance between a source at a
redshift $z_1$ and another source along the same line of sight
at a redshift $z_2$ (for $\Omega_0<1$) is
\begin{equation}
     \chi_2 -\chi_1 = \sqrt{\Omega_0 + \Omega_t -1}
     \int_{z_1}^{z_2} \, {dz \over E(z)}.
\label{chiequation}
\end{equation}
If $\Omega_t$ is chosen so that the
polar-coordinate distance of the CMB surface of last
scatter is $\chi_{LS}\simeq \pi$, then the CMB we observe comes
{}from a causally-connected patch at the antipode of the
Universe.  From Eq.~\ref{chiequation}, the condition on
$\Omega_t$ for this to occur is
\begin{equation}
     \Omega_t\simeq\left[ {\pi \sqrt{1-\Omega_0} \over {\rm arcsinh}
     ( 2 \sqrt{1-\Omega_0}/\Omega_0)} \right]^2 +1 - \Omega_0.
\label{conditioneqn}
\end{equation}
With this imposed, the texture density $\Omega_t$ increases from
1.6 to 2.5 for $\Omega_0$ between 0.1 and 1.

Is this a realistic possibility?  As we discuss below, this
model is fully consistent with our current knowledge of the
Universe.

Incidentally, one could also ``close'' a low-density Universe
with a large cosmological constant (although such a model with
the CMB at the antipode is likely ruled out by quasar-lensing
statistics$^{6)}$).
However, the Friedmann equation is altered in such a model, so
the expansion rate affects the
classical cosmological tests.  In the model discussed here,
the expansion is identical to that in an open FRW
Universe.  Therefore, quantities that depend only on the
expansion, such as the deceleration parameter, the age of the
Universe, or the distribution of quasar absorption-line redshifts,
do not probe $\Omega_t$.  Furthermore, the growth of
density perturbations is the same as in a standard open Universe, so
dynamical measurements of $\Omega_0$ (e.g., from
peculiar-velocity flows) will also be insensitive
to $\Omega_t$.  Effects due to geometry arise only at ${\cal
O}(z^3)$ since $\sin\chi$ and $\sinh\chi$ differ only at ${\cal
O}(\chi^3)$; therefore, this Universe will differ from an
open Universe only at $z\gtrsim1$.

We now turn to cosmological tests that probe the
geometry of the Universe.  Underlying these is the
angular-diameter distance between a source at a redshift $z_2$
and a redshift $z_1<z_2$,
\begin{equation}
     d_A(z_1,z_2)= { \sin(\chi_2-\chi_1) \over (1+z_2) H_0
     \sqrt{\Omega_0+\Omega_t-1}}.
\label{angulardiametereqn}
\end{equation}
The angular size of an object of proper length $l$ at
a redshift $z$ is $\theta\simeq l/d_A(0,z)$.  With $\Omega_t$
fixed by Eq.~\ref{conditioneqn}, one
finds that the angular sizes in a flat matter-dominated Universe
can be very similar to those in a low-density closed
Universe$^{4)}$.  Proper-motion distances of superluminal
jets in radio sources at large redshift may provide essentially
the same probe as do flux-redshift relations.
The difference between the angular sizes for the standard
FRW Universe and the closed model for the same value of
$\Omega_0$ is quite a bit more dramatic than the difference
between open FRW and flat $\Lambda$ models.  
It has been proposed that evolutionary effects may conceivably be
understood well enough to discriminate between open and flat
$\Lambda$ models$^{7)}$.  If so,
then the distinction between these and the closed model
will be even clearer.

In the low-density closed Universe, the differential
number of galaxies per steradian per unit redshift is,
\begin{equation}
     {dN_{\rm gal} \over dz d\Omega} = {n_0  \sin^2[\chi(z)]
     \over H_0^3 (\Omega_0 +\Omega_t-1)E(z)},
\end{equation}
where $n_0$ is the local number density of galaxies, and the number
per comoving volume is assumed to remain constant.  Again, 
one finds that the number-redshift relation for a flat
matter-dominated Universe may be mimicked by a
low-density closed Universe.

The redshift thickness $\delta z$ and angular size $\delta
\theta$ of a roughly spherical structure that grows with the
expansion of the Universe will have a ratio$^{8)}$
\begin{equation}
     {1 \over z}{\delta z \over \delta \theta}=
     {E(z)\sin[\chi(z)] \over z \sqrt{\Omega_0+\Omega_t-1}}.
\end{equation}
It turns out that this function is
significantly lower in a low-density closed Universe than
in an open Universe and in a $\Lambda$ Universe.
Curiously, it depends only very weakly on the value of
$\Omega_0$ and therefore provides an $\Omega_0$-independent
test of this closed model.  A precise measurement may be
feasible with forthcoming quasar surveys$^{9)}$.

The probability for lensing of a
source at redshift $z_s$ for $\Omega_0<1$ and
$\Omega_t+\Omega_0>1$ relative to the fiducial case of a
standard flat Universe is$^{10)}$
\begin{eqnarray}
     P_{\rm lens} & = & {15\over4} \left[ 1- {1\over
     (1+z_s)^{1/2} } \right]^{-3} \nonumber \\
         & & \times \int_{0}^{z_s}\, {(1+z)^2 \over E(z)} \left[
	 {d_A(0,z)d_A(z,z_s) \over d_A(0,z_s)} \right]^2\,dz.
\end{eqnarray}
The current observational
constraint is roughly $P_{\rm lens}\lesssim5$.  If $\Omega_t$
is chosen so that the CMB comes from the antipode, then $P_{\rm
lens}<2.5$ for $0<\Omega_0<1$.  Hence the model is consistent with
current data and is likely to remain so.

So far, we have investigated several
tests that depend on the geometry.  However, each of these also
depends on the expansion of the Universe, so no single test can
determine the geometry unless the matter density is fully
specified.  Furthermore, these involve observations at
large redshifts where observations are tricky and evolutionary
effects may be important.

So how does one determine the geometry of the Universe?  CMB
temperature maps attainable with forthcoming satellite and
interferometry experiments will likely provide the best
determination of the geometry of the Universe$^{11,12)}$.
Although the detailed shape of the
anisotropy spectrum depends on a specific model for
structure formation, it has structure
(acoustic peaks) on angular scales smaller than that
subtended by the horizon at the surface of last scatter.  This
angle depends primarily on the geometry and only weakly on other
undetermined cosmological parameters; in a standard FRW
Universe, it is
$\theta_{LS} \simeq \Omega^{1/2}\,1^\circ$, where $\Omega$ is
the {\it total} density of the Universe.  Therefore,
measurement of the location of the first acoustic peak
provides a reliable determination of the geometry of
the Universe$^{11)}$.  Furthermore, it can be shown that
with forthcoming all-sky CMB maps with sub-degree angular
resolution, $\Omega$, may be determined to better than 10\% and
perhaps as good as 1\%.$^{12)}$

Thus, for the first time since the discovery of the
expansion raised the issue, the geometry of the Universe may
finally be determined.  The location of the first acoustic peak
will therefore provide a precise
test of inflation, which predicts a flat Universe, and will test
alternative models such as the low-density closed Universe
discussed here.
Finally, what about the homogeneous matter with an energy density
which scales as $a^{-2}$?  If this is due to a topologically
stabilized scalar-field configuration, as discussed above, then
the symmetry-breaking scale must be of order the Planck scale if
$\Omega_t$ is of order unity.  Furthermore, the global symmetry
must be {\it exact}.  If confirmed, this model would therefore
have significant implications for Planck-scale physics$^{13)}$.

\bigskip

This work was supported in part by the D.O.E. under contract
DEFG02-92-ER 40699 and by NASA under contract NAG5-3091.

\end{document}